\documentclass[%
 reprint,
 amsmath,amssymb,
 aps,
]{revtex4-1}

\usepackage{url}

\usepackage{graphicx}
\usepackage{dcolumn}
\usepackage{bm}
\usepackage{siunitx}
\usepackage{physics}

\begin{document}

\preprint{APS/123-QED}

\title{Quantum Biology:\\Can we explain olfaction using quantum phenomenon?}

\author{Chukwuemeka Asogwa}
\thanks{casogwa@uwaterloo.ca}%

\affiliation{Department of Physics and Astronomy, University of Waterloo, Waterloo, Ontario, N2L 3G1, Canada 
}%




\date{\today}

\begin{abstract}
The sense of smell is an important part of living organisms. It assists with the interaction of the living organism with its environment. The mechanism by which smell is detected and identified is not fully known. Earlier, shape theory was proposed as an explanation for odor perception mechanism. This theory posits that the shape of the odorant must fit in its complementary olfactory receptor for an odor to be identified -- just like key fits in a lock. This theory turns out to have some limitations, thus leading to the proposition of new theory called the vibrational theory of olfaction. In vibrational theory, the nose is regarded as a spectroscope that detects vibration of odorants. However, this happens to be currently controversial and actively debated. In this paper, I will give a review of the history of olfaction, the challenges, and then explain some new experiments (both on human and fruit flies) that support or refute the vibrational theory of olfaction. 

 
%
\end{abstract}

\pacs{Valid PACS appear here}
\maketitle


\section{\label{sec:level1}Introduction}
\subsection{Sensing of odor by living organisms}
To understand how life began was one of the earlier challenges of scientists and philosophers. As back as the nineteenth century, a good number of scientists accepted life as a kind of magic matter which has no chemical nor physical explanation. This ideology lasted until an assumption that life has a chemical recipe. Life was then known to be a combination of two or more chemicals. This led to the suggestion of `producing life' by mixing up chemicals in the lab. This idea was a blind alley as life has not been produced using inanimate objects till date.

How scientists understand the beginning of life was revolutionized in 1944 by Erwin Schr$\ddot{o}$dinger in his published lectures; titled\textit{ What is life?}\citep{schrodinger1943life}. This famous publication gave rise to the study of molecular biology. Schr$\ddot{o}$dinger was of the opinion that there is a quantum mechanical process involved in the stable transmission of genetic information from one generation gene to another generation and that this occurs unaware of the gene encoding roles. This was backed up by other founders of quantum mechanics, Neil Bohr, Werner Heisenberg and Eugene \cite{abbott2008quantum}. As science develops, quantum mechanics explains the atomic and molecular structure component of a matter and the states it can appear to be. Life for a physics at this time was considered weird, therefore it could only be explained by a weird phenomenon like quantum mechanics \cite{mcfadden2016life}. The study of molecular biology was actively growing, arguing that the shapes of molecules and their chemical affinities compost the functionality of the cell. These properties are phenomena that quantum mechanics can explain. 

There are several proposals that the fundamental level of all things is quantum mechanical. To understand life on a higher level, some quantum concepts such as entanglement, superposition states, spooky action at a distance and quantum tunneling need to be both experimentally and theoretically proven to be connected with the processes involved in the definition of life. How energy is captured through photosynthesis and the olfaction science are suggested to be connected with some of these quantum concepts \cite{turin1996spectroscopic, malcolm1938scientific}. Among these two processes, we consider olfaction in detail. 

Olfaction is the ability to detect and distinguish different odors. These odors come as airborne molecules. Different living things adapt peculiar method in detecting and distinguishing odors. Mammals and insects, in general, can use the intensity, whether it is offensive or fragrant to remark an odor and able to identify it next time. Humans find it difficult to effectively identify odor with little or no recognizable intensity, fragrance or offensiveness. To detect where the smell is emerging from or even notice the smell is also difficult. Some animals like dogs, anemonefish, reef fish and sharks have good ability to detect the smallest strength of odor beyond human's capability. Example:
\begin{itemize}
\item Fish have a good sense of smell that helps them to locate their home even when ocean current drift them away from their home. 
\item Shark, in the same way, has the ability to detect the smell of a drop of blood as far as a kilometer away. 
\item Unlike human nostrils that are used mainly for breathing, sharks' nostrils are located on the underside of the snout and used mainly for smelling. As they swim in the ocean, water flows through their nostrils and down to the sensory cells. They have high sensitive sensory cells. The side of the nostrils the smell hits first tells the shark where the smell is coming from, with this, they hunt for the wounded prey. 
\end{itemize}

In 2007, Gabriele Gerlach \cite{gerlach2007smelling} designed an experiment to verify the theory behind larval reef fish identifying their home. She called her experimental setup a `two-channel olfactory choice flumes test'. She collected two flumes of seawater; one from reef they were hatched and the other from a far away reef. She then placed the larval reef fish downstream of the two flumes. She observed the preference of the larval reef fish between the two flumes of seawater. It came out to be that the larval reef fish preferred the seawater which they were hatched. The ability of the larval reef fish to distinguish between the two seawater was ascribed to their smelling ability. 

Similarly, Daniella Dixson \cite{dixson2008coral} 
confirmed that an anemonefish was able to distinguish between the collected water from their habitat reef and the water collected from an offshore. Michael Arvedlund \cite{arvedlund2009senses,arvedlund2009first} 
designed a similar experiment to Gabriele Gerlach's, to understand the ability of anemonefish in identifying their species. It was confirmed that anemonefish was able to distinguish their host species from others uncolonized by them. These experiments, together with earlier experiments and theories confirmed that fish follow a scent trail in identifying their home and species. 


 Atmospheric molecules undergo vibrational interaction, turbulence and migration from one point to the other just like molecules in water. Molecules in air dilute and disperse faster than the molecules in water \cite{mcfadden2016life}. Terrestrial creatures possess their own feature of olfaction. Example:
 \begin{itemize}
 \item Rabbits have 100 million scent cells which help them in identifying other rabbits and animals. This sense of smell is inherent from birth, guiding the bunnies to find their mothers' teat even when their eyes are closed. 
 \item Bears have been identified to possess the best sense of smell. It can smell 7 times better than the bloodhound, and 2,100 times better than humans \cite{mcfadden2016life}. This helps them to identify food as far as 20 miles away. In addition to finding food, they can detect mates, avoid danger coming from competing fellow bears and track the whereabouts of their cubs when lost while journeying. 
 \item Elephants, in the same way, can smell water, especially the underground water 12 miles away and still remember where they found it for future use. 
 \item Snakes have a different method of identifying a smell. They taste the air with their tongues gathering the particles using their damp surface, then transport them to Jacobson organ in the mouth where they will be identified as food or danger.  
\item Dogs, especially the bloodhound can identify and track an individual by following the smell of the butyric. Butyric acid is different for every animal and human, even for identical twins.
\item Just like in animals discussed above, smell helps human in sniffing out bad odors and their location, food using its aroma, and in communication.
\end{itemize}
This indicates that the nostrils have the capability of identifying and distinguishing several millions of odors. 
\section{Anatomy of the human nostrils}

\begin{figure}
\includegraphics[scale=.39]{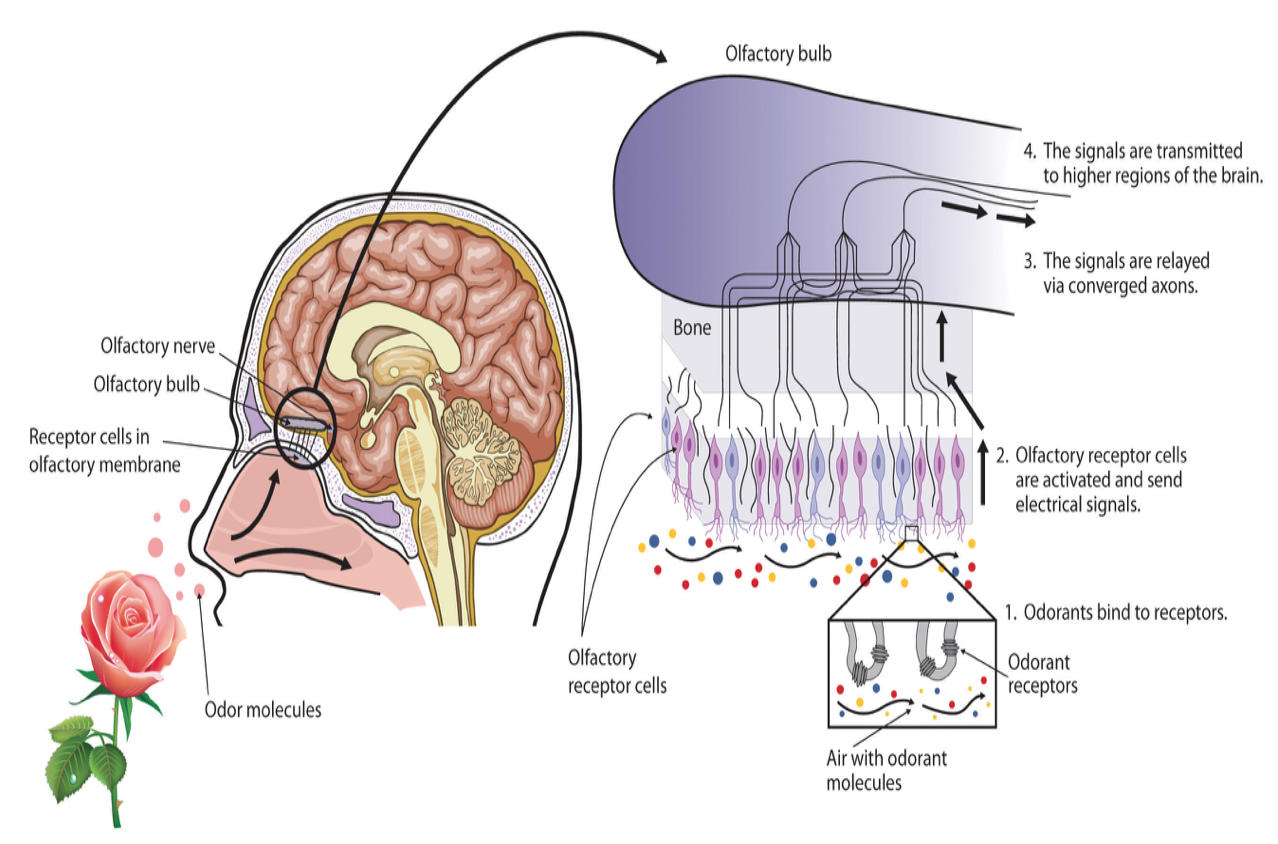}
\caption{A structure of the human olfactory system. As odor is perceived, odorant moves through the nose which acts as a channel, to meet the olfactory epithelium situated at the top of the nasal cavity. From the epithelium, it journeys to the cilia which extend from the main body of the neuron in the mucus layer. The odorant meets the mucus and the inside of the cell. Figure from \cite{olfac}.}\label{anatomy}
\end{figure}

The nose is one of the chemosensory systems which is an intrinsic feature every human possesses. Chemosensory systems (of which taste and the vomeronasal systems are inclusive) work hand-in-hand to identify a smell. They all pick up information in the form of molecules. 
Molecules can enter the olfactory system either through the nostrils normally or the back of the throat, which happens mainly when eating food. The nostrils act as a chamber that allows air pass into the nasal cavity. The nasal cavity then assists in filtering and warming the air with the mucus lining in its inner part. The odorants are then identified inside the nasal cavity as they hit the olfactory epithelium which is situated on the roof of our nasal cavity (FIG. \ref{anatomy}). A scent signal is then sent to the brain through the olfactory nerve and olfactory bulb and an interpretation is made. The olfactory epithelium harbors millions of olfactory receptors which bind up with a particular odorant. 
This helps in identifying several varieties of odors. 

Odorants are considered to have a specific shape that binds to the olfactory receptors. The receptors are activated by different molecules of the odor. There are always variations in the strength of binding of molecules to its receptor, which can affect the ability of the brain to fully interpret a smell. Our ability to detect various smells lies in the complexity of the interaction between the receptors and the odorants. This, in the actual sense, shows that the final smell we perceive is an integration of various odorants interacting with various receptors and generating encoded information \cite{malnic1999combinatorial}.

\begin{figure}
\includegraphics[scale=.45]{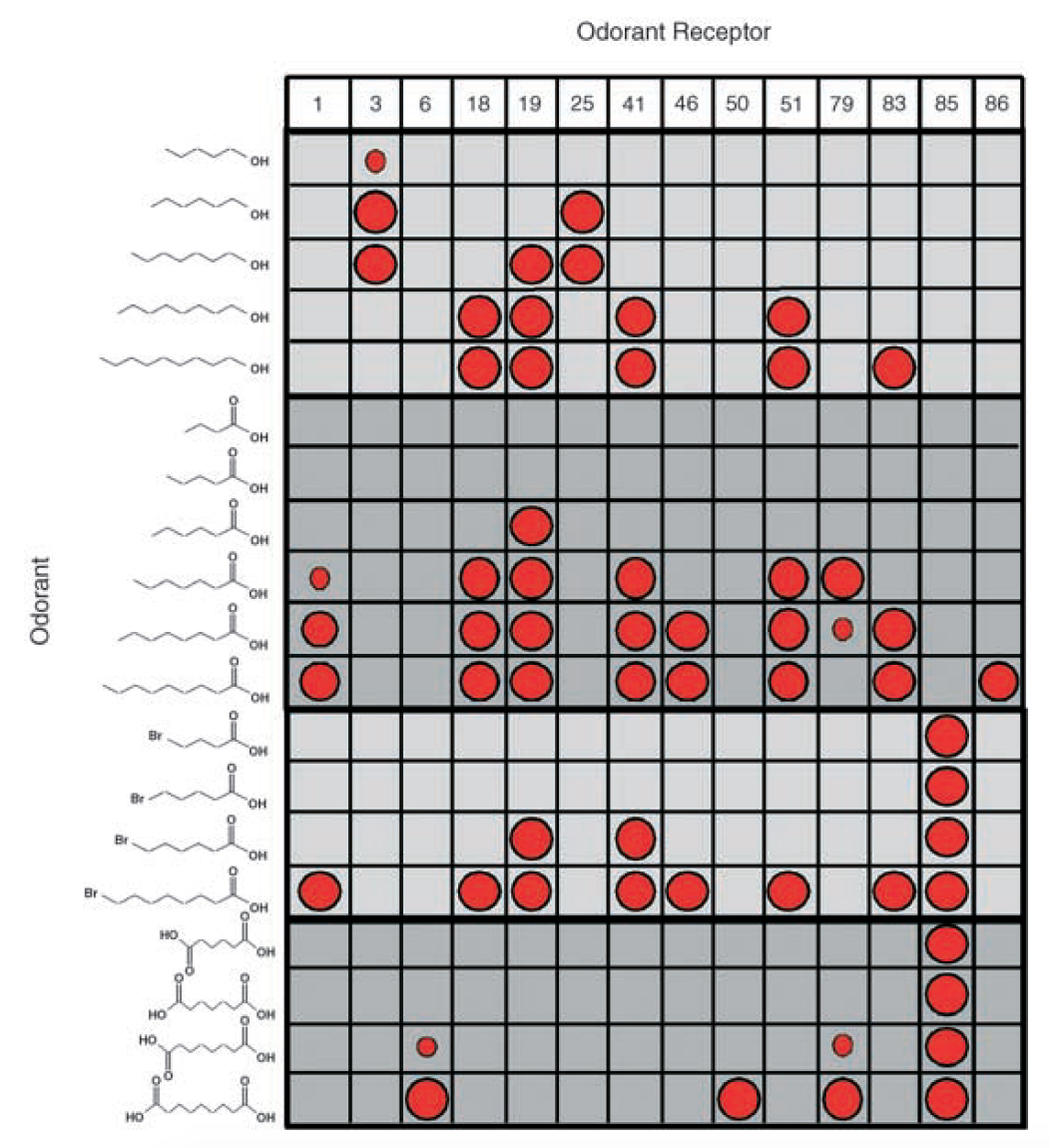}
\caption{The red circles represent `combinatorial code'. Large circles represent large responses and small circles, small responses. The addition of specific individual `codes' leads to the odorant detection. Figure from \cite{malnic1999combinatorial}.}\label{combina}
\end{figure}

In human olfactory, there are about 390 types of functional receptors which somehow `tuned' in response to the different molecular stimuli given by the odorants in order to accommodate potentially thousands of odorants \cite{harini2015computational}. This is assumed to occur in a combinatorial process, where all the different receptors have a `code' that determines the characteristics of a smell \cite{buck1991novel}. One odorant activates more than one receptor which the combination of their responses give the smell of the odorant (FIG. \ref{combina}). An odorant existing as a mixture of many odorants actives very many receptors.

The process that we are most interested in is the functionality of capturing of the odorant by the olfactory neuron. \textit{Does this process involve quantum phenomena?}

\section{Theories of olfaction}
\subsection{The `lock and key' model}

An intriguing question one would ask is; ``How does each receptor recognize its own set of odorant and binds with them?". In 1963 Amoore \cite{amoore1963stereochemical} first proposed that the response to scent works by the mechanism of a `lock and key'. The lock and key model explains that the shapes of receptor molecules and the odorants are in complementarity with each other, implying that the odorants fix into an olfactory receptor which has its shape; just like a key fits into its padlock. This idea was generated from the molecular mechanism of the behavior of enzyme \cite{tirandaz2017validity}. Another illustration of this theory is the shape fitting games that toddlers play with excitement. This involves fitting a cut-out shape into its complementary opening in a wooden or plastic board. The odorants can be imagined as the shapes trying to fit in the opening on the board, which is the olfactory molecule. 

Since odorants come in different shapes, a reasonable assumption is made that molecules having the same shape should smell alike, and odorants having different shapes should have obvious distinct smells. 

\begin{figure}
\includegraphics[scale=1]{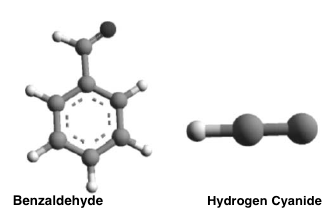}\\\line(1,0){200}

\includegraphics[scale=.4]{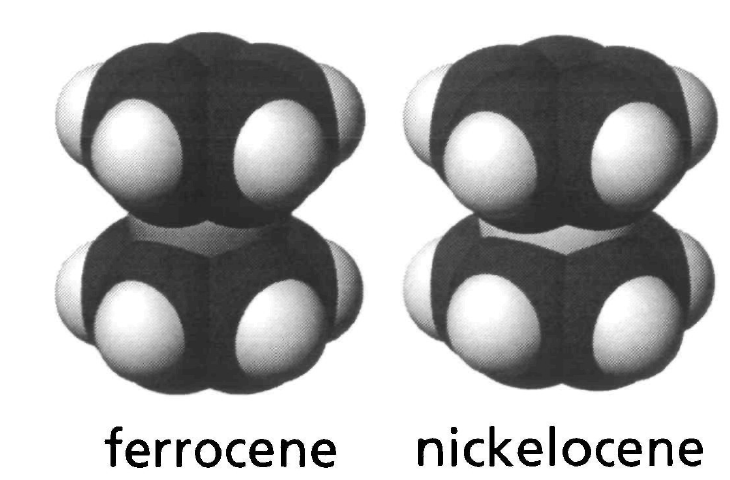}
\caption{Top: The structure of benzaldehyde and hydrogen cyanide. They both have the same smell (bitter almond) but have different structures \cite{brookes2011olfaction}. Bottom: Models of space-filling ferrocene and nickelocene. They have almost identical shapes and very different smells. Ferrocene and nickelocene have a spicy and oily/chemical smell respectively. 
Figure from \cite{turin1996spectroscopic}.}
\label{ferro}
\end{figure}

\begin{figure}
\includegraphics[scale=.5]{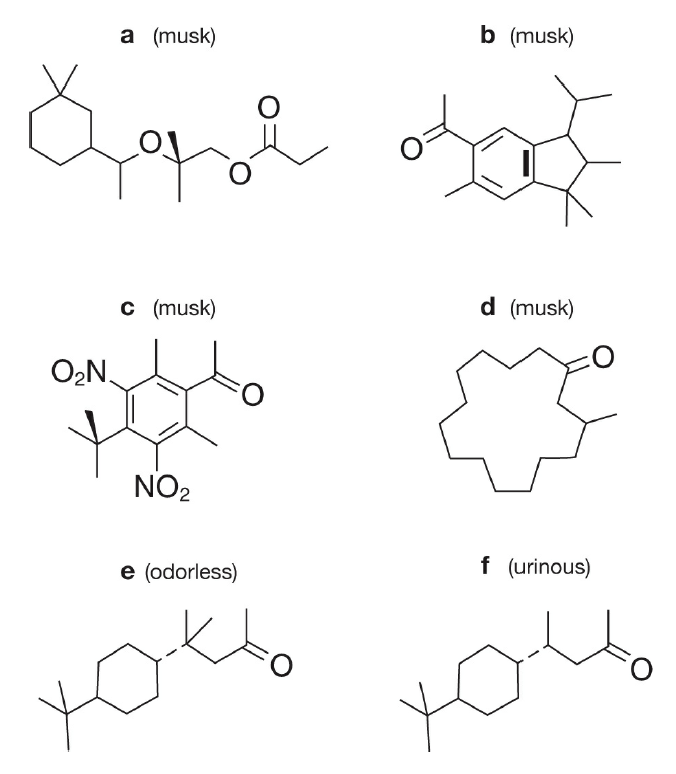}
\caption{Odorants $(a)$ to $(d)$ have different structures, but the same odor (musk), while odorants $(e)$ and $(f)$ have the same structure but different odors. Figure from \cite{mcfadden2016life}.}\label{musc}
\end{figure}

The study of the structures of different molecules has shown that this assumption does not work \cite{turin1996spectroscopic}. Notwithstanding this powerful explanation of binding of odor structurally, it has been shown that there exist molecules with different shapes yet smell alike, and compounds with similar structures yet smell differently \cite{brookes2009odour,bentley2006nose,turin2003structure}.
Odorants like ferrocene and nickelocene have similar structures 
but different odors. 
Nickelocene has a cycloalkene odor, while ferrocene smells camphoraceous. Also, hydrogen cyanide and benzaldehyde have different structures but the same odor (bitter almond). FIG. \ref{ferro} shows the space-filling model of the odorants. Other examples are illustrated in FIG. \ref{musc}. Compounds $(a)-(d)$ have different structures but smell the same (musk), while $(e)$ and $(f)$ have a similar shape and still possess distinct smells. Compound $(e)$ is identified to be odorless while $(f)$ smells like a urine. These pose a problem in `key and lock' theory. 

\subsection{Vibrational theory}

In 1938, Malcolm Dyson proposed that the nose detects the vibrational frequency of an odorant rather than its shape \cite{malcolm1938scientific}. This revolutionized the olfactory science from the traditionally accepted `key and lock' to the vibrational theory of the molecule. Dyson observed that chemicals that have the same smell are usually made up of compounds having the same chemical groups (eg., C$=$O). These chemical groups define to large extent the properties these molecules possess; smell being part of the properties. Furthermore, there exist some other chemical groups (such as \textit{thiol} (SH) group) that determine the smell of an odorant irrespective of their shape. Dyson noticed this in some chemical groups with sulphydryl (SH), having a hydrogen atom attached to a sulfur atom. This smells like rotten-egg. He measured the vibrational frequencies of the compounds following the Raman principle of light scattering \cite{raman1930molecular}.  


Light bounces off an atom when hit on surface elastically just like a ball bounces off a hard surface when hit against a hard surface. This occurs without any loss of energy. This means the energy is conserved. Chandrasekhara Venkata Raman in 1930 earned a Noble prize for his work in the field of light scattering. He noticed that the wavelength of some of a deflected light changes as it transverses a transparent material. This phenomenon was named after him as Raman scattering \cite{raman1930molecular}. This, in principle, means that light can also scatter inelastically, thereby losing some energy to the molecular bonds they hit. As the light hits the molecule, vibration occurs and the observed scattered light surfaces with less energy. This decrease in energy causes a decrease in the frequency of the photon. The amount of the lost energy gives the Raman spectrum \cite{malcolm1938scientific}. This spectrum has a specific feature called the `signature', attributed to the particular chemical bonds.

This underpins Dyson's follow-up theory that the nose could actually be seen as the `spectroscopy' that detects the vibrational signature frequencies of different chemical bonds. He noticed that there is a strong correlation of some frequencies in the Raman spectra with a particular odor \cite[for more details see][]{malcolm1938scientific}. All compounds having a terminal sulfur-hydrogen bond were identified to have a Raman frequency within the same range. 

\begin{figure}
\includegraphics[scale=0.75]{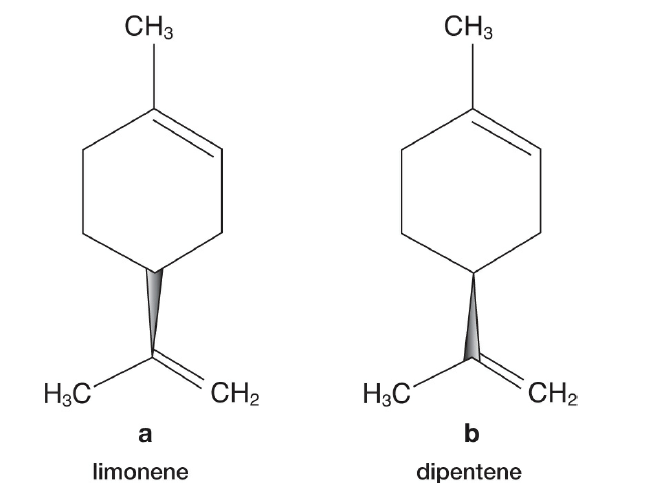}
\caption{Both limonene and its mirror-image dipentene have different smell while they are observed to have the same spectrum. Figure from \cite{mcfadden2016life}.}\label{limonen}
\end{figure}

Applying Dyson's theory to the explanation of olfaction was a difficult one. One is the methodology by which our nose acts like spectroscopy to collect the smell in form of a scattered light. Another is the involvement of light in the process. The theory became futile when it was observed that chiral molecules having the same chemical structure and identical Raman spectra could be easily distinguished by our nose. This shows that molecule can have different smell even when they have the same chemical structure and identical Raman spectra. A general example is the limonene, regarded to be right-handed molecules and its left-handed (mirror-image) molecule as dipentene (FIG. \ref{limonen}). They both have the same molecular bond, thus the same Raman spectrum but different odors. Also, carvone (one of the chemical components of seeds) and caraway (belonging to the family of flowering plant called Apiaceae) have the same Raman spectrum but different odors.

\subsection{`Weak shape' or odotope model}
\begin{figure}
\includegraphics[scale=.7]{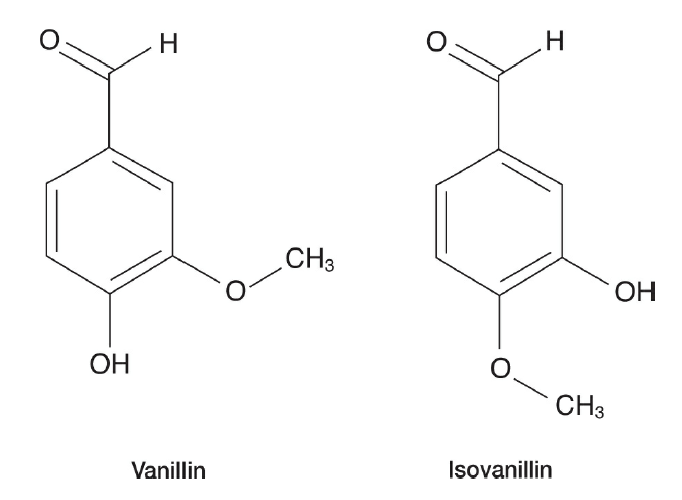}
\caption{Vanillin and isovanillin have the same chemical parts but can have different odors. Figure from \cite{mcfadden2016life}.}\label{odotope}
\end{figure}

The reason why molecules with different shapes are identified by olfactory receptors as the same odor led to the proposal of the `weak shape' or odotope theory proposed in 1994 by Kensaku Mori and Gordon Shepherd \cite{mori1994emerging}. This theory argues that there must be molecular shape recognition somewhat in odor detection. This proposes that the binding of the molecule to the olfactory receptor is based on the shape of the substructure (ie., the component chemical group) rather than the entire shape of the molecule. This theory fails for molecules with the same component of a chemical group but arranged in a different pattern. An example is vanillin and isovanillin (FIG. \ref{odotope}) having the same chemical parts but arranged differently, thus having different odors. The vibrational theory is also not yet able to solve this problem.

\subsection{Turin's theory of inelastic electron tunneling spectroscopy (IETS)} 

Based on Dyson theory of vibrational theory, Turin proposed that the principle which the olfactory receptors use to detect vibrations of chemical bonds could be related to the concept of quantum tunneling of electrons \cite{turin1996spectroscopic}. 
Quantum tunneling involves the seeping through of electrons or photons as they encounter a barrier which they do not have the energy to classically surmount. This concept of quantum tunneling of an electron was earlier developed by Henri Becquerel in 1896 while researching on radioactivity. 

\begin{figure}
\includegraphics[scale=.7]{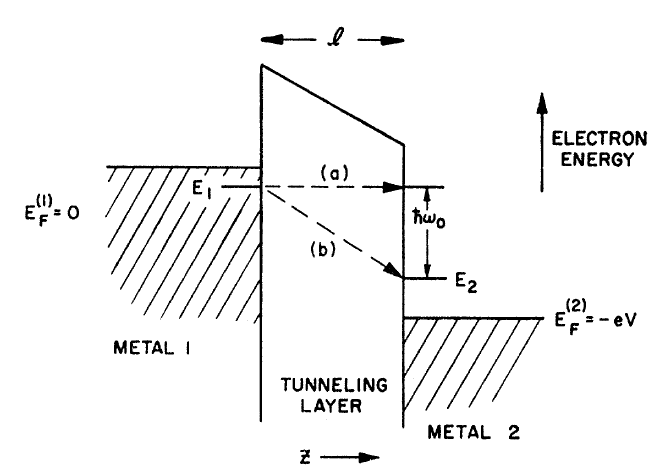}
\caption{A demonstration of electron tunneling through a metal junction. The electron can tunnel through the tunneling layer (insulating barrier) either $(a)$ elastically (ordinary tunneling) or $(b)$ inelastically. The electron only tunnels inelastically if there is a molecule present in the tunneling layer with a vibrational mode $\hbar\omega_0$, which the tunneling electron puts into excitation. Energy loss occurs for the electron to tunnel to the other metal if only $eV\geq$ $eV_0=\hbar \omega_0$, where V$_0$ is the threshold voltage. Figure from \cite{lambe1968molecular}.}
\label{fig:iest}
\end{figure}
This principle involves the quantization of the energy levels within a molecule with an arbitrary energy gap $E=\hbar \omega_0$, where $\omega$ is the resonant frequency of a particular vibrational mode and $\hbar$ is the reduced Planck constant. FIG. \ref{fig:iest} shows an illustration of the energy process involve in inelastic tunneling of an electron. There are two metal plates closely placed to each other and separated by a miniature gap (the tunneling layer). Metal 1 is negatively charged by accommodating electrons, while metal 2 is positively charged (electron acceptor). The two Fermi levels of the metals are separated by $eV$, where $V$ is the applied voltage. The Fermi energy in metal 1 $E_F^{(1)}$ is higher than in metal 2 $E_F^{(2)}$, and can be related by $E_F^{(1)}-E_F^{(2)}=eV$. Once there is an application of $V$ in the junction, electrons present in the metal 1 can classically transit horizontally to the empty state in metal 2 without losing or gaining any energy. This process of tunneling across the barrier without loss of energy is called elastic tunneling. In the actual sense, electrons can only tunnel from the metal plate to the other if there is an empty spot available at the acceptor side of the metal plate with the same energy level. The electron will happen to lose its energy if it has to enter a slot with different energy level. This process then results in an inelastic tunneling. When a molecule is present in the tunneling layer between the two metals, as
the electron tunnels across the molecule, its energy is absorbed by the molecule, and an excitation of a phonon in the molecule occurs. This creates bias voltages that couple to give the vibrational spectrum of the molecule between the two metals. 

The current-voltage characteristic $(I/V)$ of the elastic tunneling gives a linear relationship. For inelastic tunneling, the process of the tunneling of the electron inelastically results to an increase in conductance at the voltage $eV_0\simeq\hbar\omega_0$ of which a second derivative of the current-voltage characteristics $(d^2I/dV^2)$ shows clearly a defined peak in the spectra which appear like Dirac delta distribution. 
This tells the energies of the vibrational transitions of the molecule in the tunneling layer. A probe of the energy difference between the electron at the donor and the acceptor plates gives some insight on the molecular bonds of the molecule between the two metal plates. 

Following this explanation, Turin proposed that the olfactory system works with the same principle, with the olfactory receptor serving as the IETS plates, and the electron that is at the acceptor site results in the production of the G-protein molecular torpedo, which triggers the olfactory neuron to send a signal to the brain for interpretation. The explanation of the biological IETS is given below.

\subsubsection{Biological IETS}
\begin{figure}
\includegraphics[scale=.7]{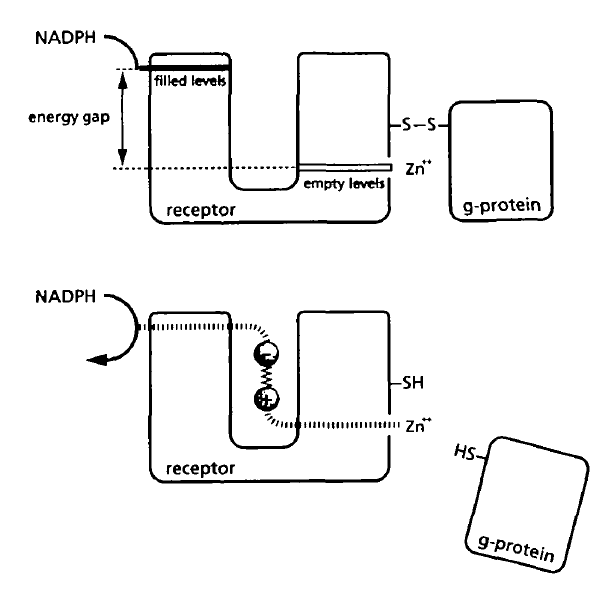}
\caption{Turin's model of IETS. An electron from NADPH is accepted at the receptor protein. The electron will not be able to tunnel when the receptor binding site is empty (as shown on top) because there is no empty level available with the right energy. This means when no odor is present to normalize the energy difference at the receptor and acceptor, no tunneling happens. On the presence of an odorant (represented by the dipole) at the binding site (as shown at the bottom), the electron tunnels losing energy during the tunneling through excitation of the odorant's vibrational mode matching the energy gap between the donor and acceptor. Electrons flow through the protein and reduce the disulphide bridge via a zinc ion thereby a release of the G-protein is triggered. Figure from \cite{turin1996spectroscopic}.}\label{turiniest}
\end{figure}

The production of electron or hole by a source that enables the flow of charge requires some energy input. It is assumed that since human cells have some voltage of order $0.5\text {V}$, then the source of the voltage is not considered a problem \cite{mohseni2014quantum}. In the biological system, for Turin's inelastic electron tunneling to occur, we require a source of the electron, removal mechanism from the electron source, the right energy levels, and a possible donor and acceptor. It is accepted that the electron transfer in biology is aided by series of oxidation and reduction reactions within several biomolecules \cite{rawson2002cell}. The transfer uses, to a very large degree, metalloproteins \cite{williams1990overview}, so the theory of IETS can be applied in the biological system. Turin \cite{turin1996spectroscopic} suggested that the source of the electron could be the nicotinamide adenine dinucleotide phosphates (NADPH). NADPH binds to the electron donor at one side of the gap through several amino-acid motifs which build up the tertiary structure of a protein \cite{lodish1995molecular}, and at the other side of the gap, an acceptor having zinc-allegedly coming from the anosmic attributed to zinc deficiency in the diet, replenished through supplementary and dietaries \cite{brookes2009microscopic}. The zinc anchors the olfactory G-protein as shown in Turin's proposal in FIG. \ref{turiniest}.
The G-protein is release as a result of the oxidation of disulphide bridge as the electron tunnels to the acceptor. This suggests the pathway for olfactory signal transduction in which odors are bind to specific receptors and activate specific G-proteins, and then transduce an intracellular signal causing the activation of second messenger systems for possible interpretation by the brain \cite{buck1991novel,rawson2002cell}. The vibrational frequency of odorant could be measured via regarding the receptor protein to act as a spectrometer designed to detect a particular quantized vibration that relates to the difference in energy. Turin posits that there are about 10 receptors which get tuned during this process of electron transfer, and the signal produced is a combination of overlapping messages from different cells involved in the process. Olfactory cells have been found \cite{restrepo1993human,thurauf1996cyclic} (using the whole-cell patch Clamp technique \cite{hamill1981improved}) to have a resting potential in the range of $(-50$ to $-65\si{mV})\pm 12\si{mV}$ and with average membrane capacitance of $3.9\pm \si{1pF}$. A biological IETS does not involve a scan over a range of frequencies like the conventional IETS does, rather it involves a build-up of spectrum piecewise by a series of receptors tuned to a range of different frequencies \cite{turin1996spectroscopic}. The frequency range is limited only by the emf from the NAPDH at the receptor site. Most observed molecular bonds in odorants have been calculated \cite{turin1996spectroscopic} to have vibrational frequencies ranging from 0 to $4,000 \si{ cm}^{-1}$ and also a corresponding energy range. Obtaining the actual measurement of the energy transfer associated with every receptor would help in the classification and grouping of odorants according to the observed energy. This implies that several receptors would be involved in order to cover the vibrational spectrum and each can be tuned to a particular range of the vibrational spectrum. The acceptor and the donor energy levels are prone to thermal broadening of the range $2\si{kT} (\sim400\si{cm}^{-1})$ implying that the spectrometer of the nose can have poor resolution \cite{turin2003structure}. Thus about 10 receptor can cover receptor types within the range of $\si{0-400cm}^{-1}$. This arrangement is synonymous to that found in other spectral senses like vision and hearing where the broadly tuned receptor groups lie within the complete spectrum.

However, electron transfer does not happen at all if the donor is not filled with the specific energy level and the acceptor having empty energy levels just like in ordinary IETS.
\subsubsection{IETS spectra calculation}
\begin{figure}
\includegraphics[scale=.479]{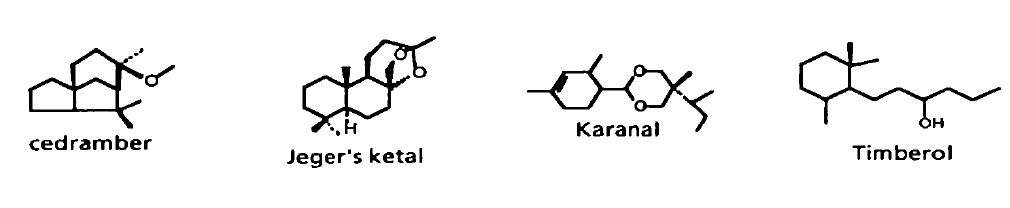}
\includegraphics[scale=.4]{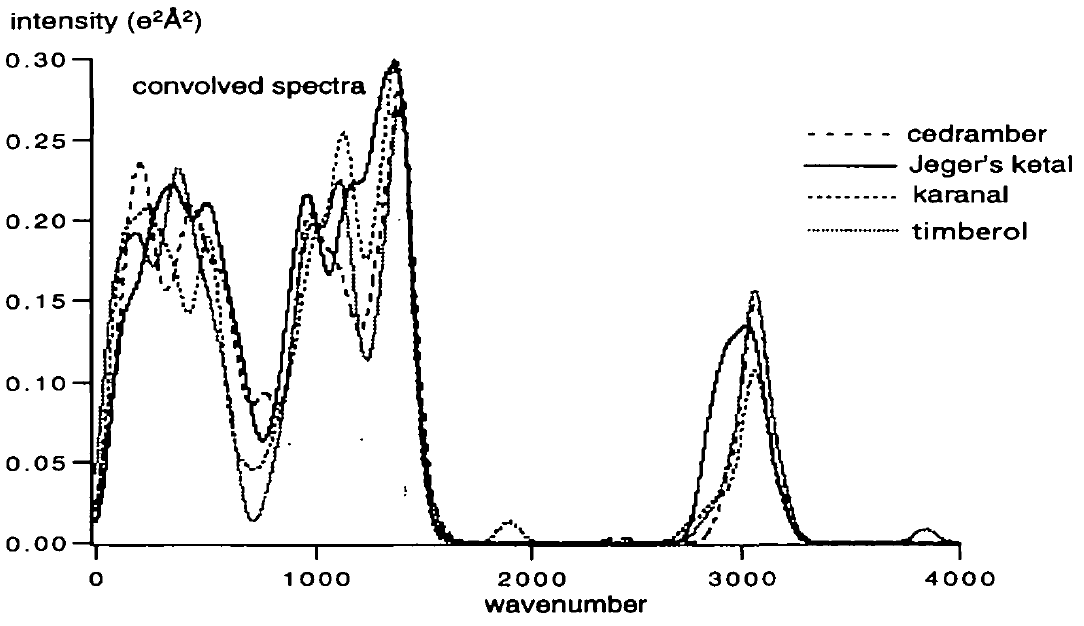}
\caption{Top: The structure of ambergris odorants having similar odors but different structures. The similar odors were identified to have obvious similar CHYPRE spectra despite having different structures. Figure from \cite{turin1996spectroscopic}.}\label{ambergris}
\end{figure}

\begin{figure}
\includegraphics[scale=.42]{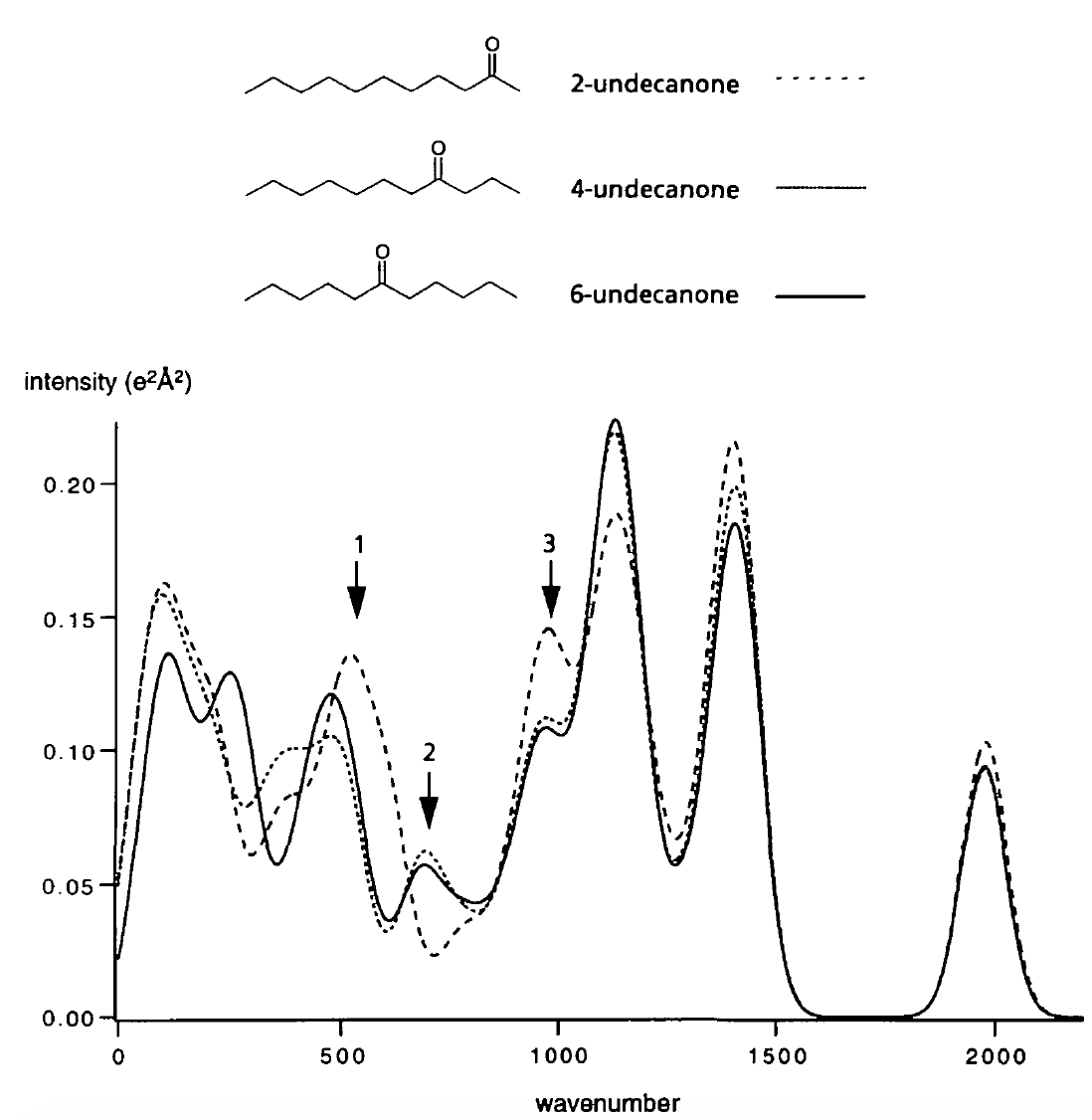}
\caption{Top: The structure of three undecanones differing in the position of the carbonyl group. As a result of the shift in the position of the carbonyl from 4 to 6, the peaks of the intensities (indicated by arrow) are affected. The spectra of 2- and 6-undecanone are clearly different while 4-undecanone has an intermediate spectrum between the two, except within the region of $5-600$cm$^{-1}$ wavenumber where it has a close match with that of 2-undecanone. Figure from \cite{turin1996spectroscopic}.}\label{undecanone}
\end{figure}

If electron tunneling is the mechanism of odor detection, then it implies that there should be a correlation between the tunneling spectra of the molecules measured by the biological detector and their odors. The comparison of spectra from different odorants is based on 1) the frequency of a given vibrational mode, 2) its intensity recorded by the sensor and 3) the resolution of the detector system. The theory and calculation for the measurement of IETS spectra of different compounds in metal-insulator-metal junctions have been developed. The application of the theory to the biological system which involves odorant is difficult because the odorants are prone to evaporating during the vacuum deposition processes involved in the manufacturing of the tunneling junctions. Turin \cite{turin1996spectroscopic} developed an algorithm called CHYPRE (CHaracter PREdiction) which calculates the biological IETS spectra. Apparently, the difference between the calculated value from IR and the CHYPRE is that the region of fingerprint below 1500cm$^{-1}$ that is generally used as a guild to determining molecular structure is empathized more by CHYPRE than by IR. CHYPRE algorithm was then used to examine the postulate that smell of compounds can be predicted from their vibrational spectra notwithstanding the characteristic of their structures. As an example, a comparison of the spectra of the chemically related ambergris odorants having different structures and similar odor was done. The ambergris odorant used were cedramber, karanal, jeger's ketal and timberol (FIG. \ref{ambergris}). The observed convolved vibrational spectra of the ambergris odorants were seen to be very similar, in consistency with their similar odor.
Conversely, molecules having very similar structures and different odors were observed to have different spectra. As an example, the spectra of three undecanones that differ in the position of the carbonyl group were calculated. 2-Undecanone has the odor of ruewort, 6-undecanone has a fruity smell, while 4-undecanone has a smell that is intermediate between the two undecanones (FIG. \ref{undecanone}). Notwithstanding that the structures of the odorants are closely similar, 2- and 6-undecanone were identified to possess obvious different spectra.

These two examples suggest that the olfactory system uses vibrational transduction mechanism in discriminating molecules that are closely related.

\subsection{Swipe-card model}

Swipe-card model is a hybrid of the theory of `lock and key' and the vibrational theory. This model was first proposed by Marshall Stoneham and published as Brookes et al. \cite{brookes2007could}. Stoneham and colleagues argued that the shape of the olfactory receptor and the bond vibrations of the odorant give rise to what we identify as a smell \cite{brookes2007could}. The binding part of the olfactory receptor is said to function just like the swipe-card machine.

A swipe-card has a magnetic strip with encoded information on it. It can only work in a swipe-card machine if the swipe-card has the same shape, thickness, placed in the right position and has the information already in the system. 
For chiral molecules, the odorant with right or left chiral fits into the right or left olfactory receptor and can only be identified if the olfactory receptor is able to recognize it. This, in general, means that the odorant first fits into its complementary receptor shape, then triggers the vibration-induced electron tunneling. The right-handed receptor detects the right-handed molecule and gives its smell which differs from corresponding left-handed molecule detected by the left-handed receptor. 

The point still unclear is how the shape of the olfactory receptor looks, how both the acceptor and donor molecules are positioned, and the binding of the right-handed and left-handed molecules to either same or different receptor. 
\section{Electron tunneling rate}
Electron transfer in biological system involves oxidation (reduction) of certain species $(X)$ in the cell fluid \cite{rawson2002cell}, though unclear on the particular biological origin \cite{brookes2007could}. The time interval $\tau_X$ involved in the diffusion of the electron through an aqueous medium can be estimated using the standard method for computing diffusion of material through a solution from the diffusion equation and the Stokes-Einstein relation for the diffusion coefficient \cite{atkins2002equilibrium}. This gives an estimate of diffusion time as 
\begin{equation}
\tau_X=\frac{3\eta}{2n_Xk_\beta T},
\end{equation} 
where $\eta$ is the viscosity of water $(0.89\times10^{-3}\text{kgm}^{-1}\text{s}^{-1})$, $n_X$ is the concentration of $X$, $k_\beta$ is the Boltzmann constant and $T$ is the temperature (in Kelvin). The nature of $X$ or the receptor does not contribute to the result of the estimate. It has been estimated that $n_X$ will lie within $1\mu \text {M}$ to $100\mu \text {M}$, thus a substitution of typical biological system in the equation results to a value of $\eta_X$ within the range $0.01$ to $1\text {ms}$ \cite{brookes2007could}. The crossing process of the electron through the odorant to the odorant receptor can be described using Marcus theory \cite{marcus1964chemical}, and the rate in which the electron tunnels can be approximated using Fermi's Golden Rule:
\begin{eqnarray}
\frac{1}{\tau_{i-f}}=\frac{2\pi}{\hbar}|\ \left<\psi_f|\hat{H}|\psi_i\right>|^2\rho,
\end{eqnarray} 
where $\psi_i$ represents the eigenfunction of the initial eigenstate, $\psi_f$ is the eigenfunction of the final state and the density of the final state or the Franck-Condon (FC) factor is $\rho$. $\hat{H}$ is the Hamiltonian tunneling matrix.



Eyring-Polanyi equation \cite{eyring1935activated} is applied to describe the effect of temperature on the rate of chemical reaction. The transitional rate constant of an electron from donor to acceptor will be given as \cite{brookes2017quantum}
\begin{equation}
k=\kappa B\left(-\frac{\Delta G^\ddagger}{k_\beta T}\right),\label{tunelrate}
\end{equation} 
where the electron collision frequency $B$ depends on the phase of the reacting molecule. For bimolecular reaction, $B$ becomes the liquid phase collision frequency, $B\sim 10^{11}\si{M}^{-1}\si{s}^{-1}$, but for monomolecular reaction (intra molecular reaction), $B$ becomes a vibrational frequency $1/(\beta \hbar)$, $B\sim10^{13}\si{s}^{-1}$ \cite{devault1984quantum}. $\Delta G^\ddagger$ is the Gibbs energy of activation, $k_\beta$ is the Boltzmann's constant and $T$ is the absolute temperature. $\kappa$ is associated to the electron-transfer matrix element and gives the probabilistic value of the electron-transfer in the reaction. It determines whether the reaction is adiabatic or non-adiabatic. For $\kappa=1$, the adiabatic reaction dominates, while for $\kappa<1$, the non-adiabatic regime occurs. At the classical regime, $\kappa\rightarrow|H_{DA}|^2$. Using the Gibbs energy, $\Delta G^\circ=\Delta H-T\Delta S^\circ$, equation \ref{tunelrate} can be written as
\begin{equation}
k=\kappa B \exp\left(\frac{\Delta S^\ddagger}{k_\beta}\right)\exp\left(-\frac{\Delta H^\ddagger}{k_\beta T}\right),\label{rate}
\end{equation}
where $\Delta S^\ddagger$ is the entropy of activation and $\Delta H^\ddagger$ is the enthalpy of activation. Comparison of equation \ref{rate} with the Arrheminus equation given as \cite{devault1980quantum} 
\begin{equation}
k=\kappa _\infty \exp\left(-\Delta E^\ddagger/k_\beta T\right)
\end{equation}
where $\Delta E^\ddagger$ represents the energy of activation, implies that $\Delta E^\ddagger$ corresponds to $\Delta H^\ddagger$ and $k_\infty$ corresponds to $\kappa B \exp\left(\Delta S^\ddagger/k_\beta\right)$.

\begin{figure}
\includegraphics[scale=1.2]{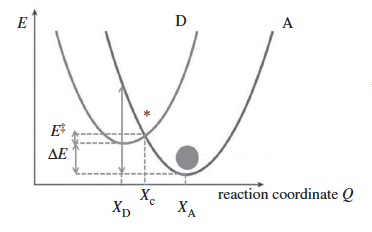}
\caption{A configuration diagram showing the total potential energy as a function of reaction coordinate $Q$. The activation energy barrier $E^\ddagger$, which separates the reactant and the product states (respectively donor D, and acceptor A) is modeled as an adiabatic reaction with $\kappa=1$. The energy difference between D and A at coordinates $X_D$ and $X_A$ is represented as $\Delta E$. $X_c$ represents the crossing point for the transiting particle at the asterisk. 
Figure from \cite{brookes2017quantum}.}\label{tunel}
\end{figure}

The configuration coordinate diagram (FIG. \ref{tunel}) shows the terms. Considering a parabolic geometry, $E^\ddagger$ can be expressed as \cite{brookes2017quantum}
\begin{equation}
E^\ddagger=\frac{\left(\lambda+\Delta E\right)^2}{4\lambda},
\end{equation}
where $\lambda$ is the reorganization energy (also called relaxation energy) coupled to the electron transfer and $\Delta E$ is the energy difference between the donor and the acceptor. $\lambda$ is a combination of the reorganization energies due to the inner shell atoms $\lambda_i$ and the surrounding solvent molecules $\lambda_o$. Combining the parameters from the inner shell vibrational modes, $\lambda_i$ can be calculated as
\begin{equation}
\lambda_i=\frac{1}{2}\sum_{j}k_{ij}Q^2_j, \quad\text{(for normal modes).}
\end{equation}
$k_{ij}$ is the Hook's law force constant and $Q_j$ represents all the harmonic oscillators arising from the displacement of the equilibrium position of the vibrational coordinates due to the electron transfer.        

The reorganization energy from the surrounding solvent molecules $\lambda_o$ is calculated by assuming the solvent as a continuous polar medium. This gives $\lambda_o$ as \cite{devault1984quantum}
\begin{equation}
\lambda_o=\frac{(\Delta e)}{4\pi\epsilon_0}\left( \frac{1}{2r_1}+\frac{1}{2r_2}-\frac{1}{r_{12}} \right)\left(\frac{1}{D_{op}}-\frac{1}{D_s} \right),
\end{equation}
where $\Delta e$ is the electron transfered from D to A, $r_1$ and $r_2$ are the radii of the two reactants, $r_{12}$ is the radius of the reactant in contact, $D_{op}$ is the square of the refractive index of reaction \cite{devault1984quantum}, $D_s$ is the static dielectric constant and $\epsilon_0$ is the vacuum permitivity. $\lambda$ can in general be approximated as \cite{brookes2017quantum}
\begin{equation}
\lambda=\frac{(\mu \omega^2Q^2)}{2},
\end{equation}
where $\mu$ represents the reduced mass, $Q$ is the normal mode, and the angular frequency of the harmonic oscillation $\omega=2\pi \nu$.

The probability of an electron to penetrate barrier during tunneling is small for inelastic channel and greater for elastic channel. From quantum mechanics, the width of the barrier contributes highly to the prediction of the probability of penetration. The probability for a particle to penetrate a square barrier is given as \cite{gray2005long}
\begin{equation}
P\propto e^{(-2/\hbar)}\sqrt{2m\Delta E^\ddagger}r.
\end{equation}

The probability varies as an exponential decay as the barrier width $r$ increases and the decay constant varies as the square root of the product of the barrier height $E^\ddagger$ and the particle's mass $m$.

To estimate the electron tunneling rate in a biological system, a generalized formula of non-adiabatic semi-classical Marcus theory describing the rate of olfaction can be obtained as \cite{brookes2011olfaction}
\begin{equation}
\frac{1}{\tau _{D_{,0}\rightarrow A_{,n}}}=\frac{2\pi}{\hbar}t^2\frac{\sigma_n}{\sqrt{4\pi \lambda k_\beta T}}\exp\left(  -\frac{ (E_n-\lambda)^2}{4\lambda k_\beta T}\right),
\end{equation}
where $E_n=\epsilon_D-\epsilon_A-n\hbar\omega_0$, $\epsilon_D$ and $\epsilon_A$ are the energy states of the donor and acceptor, and $\hbar\omega_0$ is the vibrational mode of the odorant. The factor $n$ represents the number of phonon excitation on the odorant. $n=1$ means one phonon excitation and $n=0$ means zero phonon excitation. $\sigma_n$ is poisson expression for the dependence on Huang-Rhys factor $S$, $\sigma_n=\exp(-S)S^n/n!$, $\lambda$ is the environment reorganization energy. $\lambda=\sum_{q}S_q\hbar\omega_q$, where $S_q$ is the Huang-Rhys factor for all oscillations in the environment. $t$ contains an electronic coupling matrix element that determines the strength between $D$ and $A$.

Substituting values for a typical biological system, Brookes et al. \citep{brookes2007could} found the times characterizing elastic $(\tau_{T_0})$ and the inelastic $(\tau_{T_1})$ electron tunneling from donor to acceptor to be different. Their result gives $\tau_{T_0}\sim87$ns and $\tau_{T_1}\sim1.3$ns, thus implies $\tau_{T_1}\ll\tau_{T_0}$. This indicates that the inelastic electron tunneling could be the mechanism involved in the detection of odorants by the human nose. The result of the calculation also supports that the process of the detection of odorants happens within the order of milliseconds. The total time estimate includes all the time from different stages in the process of detection of the odorant. The time from the different stages include:
\begin{itemize}
\item The time it takes the electron source to diffuse to the donor site (receptor protein).
\item The time it takes the electron (moving from the source) to be accepted at the donor site.
\item The time it takes the electron to either tunnel elastically or inelastically across the tunneling barrier.
\item The time it takes the electron to finally move from the acceptor site.
\end{itemize}

\section{A model of chiral recognition in olfaction}
One of the shortcomings of the vibrational-based model is the detection of enantiomers chiral odorants as having the same spectra and different smells. It is difficult to use a simple model of vibrational olfaction to give an explanation to this \cite{malcolm1938scientific}. Turin's explanation to this states that as a result of the different geometry of the enantiomers, the olfactory receptor seems not to be detecting some of the vibrational modes. For the case of carvone, the carbonyl in one of the two enantiomers is detected lesser intensely than the other as a result of wrong orientation of the molecule. The C$=$O group in (S)-carvone is silent. To solve this difference in smell, he suggested that `adding back' some carbonyl stretch frequency back to the (R)-carvone would shift odor character from mint to caraway. This can be achieved through the mixture of (R)-carvone with a molecule (eg., 2-pentanone) having C$=$O stretch as its dominated vibrational spectrum. In fact a mixture of 2-pentanone and (R)-carvone results to a change of its mint smell to caraway smell.

Arash Tirandaz et al. \cite{tirandaz2015dissipative} in their paper presented a quantum model of olfaction for chiral recognition. This model first checks the physical viability of odorant-mediated inelastic electron tunneling in olfactory science. Unlike the vibrational model of olfaction which is represented by a simple harmonic oscillator, the molecules are assumed to be undergoing a contorsion vibration as they oscillate between double-well potential energy surface. The contorsional mode gives the representation of the chiral recognition. To incorporate all the biological effects happening in the environment as one perceives smell, a set of harmonic oscillators stands as a representation of the biological environment. The model takes into account three main components: 1) the chiral odorant with Hamiltonian $\hat{H}_{od}$, 2) the electron that tunnels through the odorant to the receptor with Hamiltonian $\hat{H}_e$ and 3) the surrounding environment with Hamiltonian $\hat{H}_{Ev}$, all sum up to contribute to the total Hamiltonian of the system, given by 
\begin{equation}
\hat{H}_0=\hat{H}_{od}+\hat{H}_e+\hat{H}_{Ev}.
\end{equation}

Due to the chiral and the fundamental parity-violating interactions, asymmetry was introduced to account for the interactions of the molecules in the environment and the odorant. The odorant which is considered as an asymmetric double-well potential has left- and right-handed states $\ket{L}$ and $\ket{R}$ of the minima potential. As the electron tunnels through the barrier, the handed states are inter-converted. The Hamiltonian of the odorant includes the effect of the handed states. This is represented as \cite{tirandaz2015dissipative}
\begin{equation}
\hat{H}_{od}=-\frac{\omega_z}{2}\hat{\sigma}_z-\frac{\omega_x}{2}\hat{\sigma}_x,
\end{equation} 
where $\hat{\sigma}_i$ is the $i$-component of Pauli operator, $\omega_z$ and $\omega_x$ are the asymmetry and the tunneling frequencies respectively. The tunneling electron from the donor state $\ket{D}$ to the acceptor state $\ket{A}$ has energy $\epsilon_D$ and $\epsilon_A$ for donor and acceptor states respectively. These together gives a Hamiltonian of the electron as $\hat{H}_e=\epsilon_A\ket{A}\bra{A}+\epsilon_D\ket{D}\bra{D}$. The Hamiltonian of the biological environment is given as $\hat{H}_{Ev} =\sum_i\limits \omega_i\hat{b}_i^\dagger\hat{b}_i$ where $\omega_i$ is the $i$-frequency in the environment, and the creation and annihilation operators are $\hat{b}_i^\dagger$ and $\hat{b}_i$ respectively. In a broader view, the total Hamiltonian of the interaction is a combination of the three individual components, which are: 1) interaction between the donor and acceptor of the receptor with tunneling strength $\Delta$, 2) the interaction between donor and the odorant, and acceptor and the odorant, with $\gamma_D$ and $\gamma_A$ as the coupling strengths respectively, and lastly, 3) interaction between the donor (acceptor) and the environment's $i$-th harmonic oscillator having coupling strength $\gamma_{iD}$ $(\gamma_{iA})$. The interaction Hamiltonian is given as \cite{tirandaz2015dissipative}
\begin{eqnarray}
\hat{H}_{int}&=&\Delta\left(\ket{A}\bra{D}+\ket{D}\bra{A}\right)\nonumber\\\nonumber
&+&\left(\gamma_D\ket{D}\bra{D}+\gamma_A\ket{A}\bra{A}\right)\hat{\sigma}_x\\
&+&\sum_i\left(\gamma_{iD}\ket{D}\bra{D}+\gamma_{iA}\ket{A}\bra{A}\right)(\hat{b}^\dagger_i+\hat{b}_i).
\end{eqnarray}
\begin{figure}
\includegraphics[scale=.55]{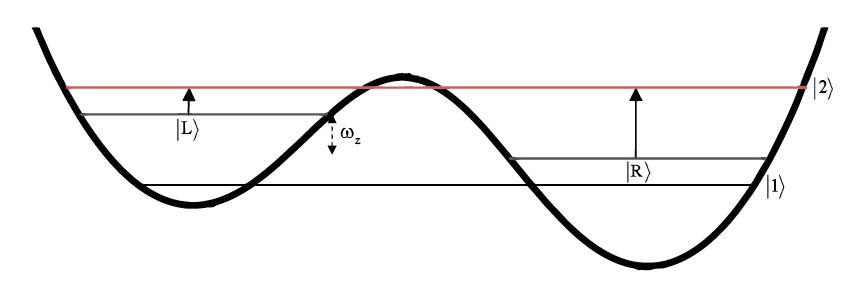}
\caption{The transitions in chiral odorant. There is no transition for elastic electron tunneling. For the inelastic electron tunneling, there is a vibrational transition from left- and right-handed states to the first excited energy state as the electron tunnels inelastically through the respective left- and right-handed enantiomers. Figure from \cite{tirandaz2015dissipative}.}\label{tunneling}
\end{figure}

The combination of the total Hamiltonian of the system gives the evolution used in calculating the electron tunneling rates. Tirandaz et al. \cite{tirandaz2015dissipative} proposed measurable parameters -- temperature and pressure which can be used to distinguish between elastic and inelastic tunneling through the potential. FIG. \ref{tunneling} shows the transition of the chiral odorant for an inelastic electron tunneling. The result from Tirandaz et al. shows that there are thresholds for which the olfactory system would be able to recognize an odor. In their calculation, they obtained different rates for elastic and inelastic tunneling. The calculated ratio of the inelastic electron tunneling rate for the left-handed enantiomers to the right-handed counterparts was found to increase with the ratio of the tunneling frequency to the asymmetry frequency. Arash and co-workers \cite{tirandaz2015dissipative,tirandaz2017validity} proposed that there is an energy difference which occurs as a result of the interactions of chiral between the donor and the acceptor. The energy difference between two enantiomers of a chiral odorant is the main factor for recognizing chirals.


\section{Experimental results on vibrational theory of olfaction}
There are various experiments conducted both on mammals and insects to confirm the vibrational theory of olfaction. One of the methods of the experiments is the isotope exchange; a replacement of hydrogen (mass 1.007) with deuterium (mass 2.014). Deuteration does not change molecular shape, atom size or bond length or stiffness, rather the vibrational modes of the odorant as a result of doubling the hydrogen mass. This results to different smells of the isotopomers. Impurities can arise in the process of preparing the deuterium of the odorant. The presence of an impurity in an odorant can affect the perception odor. To avoid a bias interpretation of smell, adequate precautions are taken while preparing the synthesis of the odorant and the final molecule collected through efficient method (eg., Gas chromatography) that would only give a pure mono-molecule odorant. 

The olfactory system of mammals and insects have certain common features as well as some unrelated features. For example, in mammals, the olfactory receptors which are proteins indicated in the cell membrane of olfactory receptor neurons (ORNs) are located in the nasal cavity, whereas in insects, they are located within sensilla pores on the antennas \cite{horsfield2017molecular}. The odorant receptors in insects are genetically different from those possessed by vertebrates \cite{kaupp2010olfactory}. Also, the number of glomeruli (a spherical structure located in the olfactory bulb of the brain where synapses happen) are different in insects and vertebrates. Fruitfly has about 62 glomeruli, 165 in the honeybee, 1800 in mice and about 1100 to 1200 in human \cite{horsfield2017molecular}. 

An intuitive question that always arises is whether both the insects and the mammals use the same mechanism to identify smell despite having some differences in their olfactory system. One would not finally conclude that both use the same mechanism or not without the full knowledge of some suspected phenomena (still unclear) that could determine odor detection. A knowledge of the structures of olfactory receptors of both insect and mammal at the atomic level would also help in the conclusion. Nevertheless, there is a common fundamental mechanism followed in distinguishing different odors, so it wouldn't be surprising when the final answer turns out to be the same mechanism. In fact, a test on anosmic \textit{Drosophila} (explained herein) has strongly suggested that flies use olfaction just like mammals to detect an odor.

The idea that olfaction is related to the vibrational frequency of odorant is still regarded as speculative since no research is yet to give the structures of the odorant receptors, the binding sites or the processes involved in the activation and binding of odorant to the odorant receptors \cite{block2015implausibility,muthyala2017testing}. 

Both humans and insects have been confirmed to have the ability to distinguish between different isotopomers. Eric et al. \cite{block2015implausibility} in their research tested the response of human musk-recognizing receptor OR5AN1, and also the mouse (methylthio) methanethiol-recognizing receptor, MOR244-3 to 1) deuterated, 2) nondeuterated and 3) $^{13}$C isotopomers. In doing so, they considered the effects of impurities and isotope effects in the interpretation of the odor perceived since some experimental results in olfaction have been criticized of having impurities. Example, an experiment by Haffenden et al. \cite{haffenden2001investigation} using benzaldehyde-d$_6$ and benzaldehyde gave that both isotopomers have statistically significant different odors supporting the vibrational theory. This result was criticized of not accounting for the perireceptor events (ie., the enzyme-mediated biochemical conversion of odorants in the nasal mucus before reaching the olfactory receptor) and not having double-blind controls to get rid of bias in their duo-trio test \cite{keller2004psychophysical}.

Indeed, there are several conflicting results on whether a human can distinguish \cite{turin1997nose,turin2003structure,haffenden2001investigation} between deuterated odorants (benzaldehyde and acetophenone) or not \cite{gane2013molecular,keller2004psychophysical}. It appears that musk isotopomers are accepted to be easily distinguished by human \cite{gane2013molecular}. Also, studies show that \textit{Drosophila melanogaster} (fruit flies) can distinguish between isotopomers of acetophenone \cite{franco2011reply,bittner2012quantum}. Likewise, training \textit{Apis mellifera} (the honey bee) makes them able to distinguish pairs of isotopomer \cite{gronenberg2014honeybees}. Though concerns have been made on the response of the \textit{Drosophila} to be behavioral and not related to olfactory receptors, and since signaling of the olfactory receptors in \textit{Drosophila} is different to the human's, then the result of the test on \textit{Drosophila} should not be finally attributed to the ability of the human to differentiate isotopomers \cite{franco2011reply,sell2014chemistry}.

To this end, instead of relying on the behavioral tests, Block et al. \cite{block2015implausibility} deemed it to be essential to test the vibrational theory of olfaction at a molecular level using receptor-based assays. They found that human musk receptor, OR5AN1 was actually able to distinguish between muscone and muscone-d$_{30}$ suggesting that the vibration theory does not apply to it neither does it apply to the mouse thiol receptor MOR244-3 or other olfactory receptors tested. Their theoretical analysis of vibrational theory also suggests that it is unrealistic in the biological content. 

However, it would be interesting to look into details of one experiment each supporting or refuting the vibrational theory of olfaction. It is worth noting that the choice of the experiments we will be focusing on was based on my behavioral response (interest) on the methods used in the experiments. We, therefore, discuss the experiments below.

\subsection{Experimental support of the vibrational theory of olfaction}
To test whether animals can distinguish an odor or not, a behavior action in responding to a given odor is examined. Franco et al. \cite{franco2011reply} examined this in their experiment using a T-maze shape olfactomers and fruit flies placed between the arms of the T-maze having odorants. The preferential response of the flies was observed by counting the number of flies that moved to each arm of the T-maze in preference to the odorant placed in there.

Acetophenone (ACP, C$_8$H$_8$O) and its deuterium atoms three, five and eight (d$_3$, d$_5$, d$_8$) were used in the first stage of the experiment. It is expected that the flies should respond differently to ACP and deuterated ACP due to their different smells and vibrational frequencies. The odorants are $>99\%$ impurity free. ACP was diluted in nonvolatile odorless isopropyl myristate (IPM, C$_{17}$H$_{34}$O$_2$) and flies placed at the T-maze arm to choose between ACP and IPM. The flies ($>15\%$ excess flies in the preferred arm) exhibited a vivid natural preference for ACP. When d$_3$-, d$_5$- and d$_8$-ACP were tested, a contrary outcome was obtained. It was found that the flies start to show preference towards the IPM. A clear preference about $15\%$ excess flies in the preferred arm IPM was seen when d$_8$-ACP was used. They also examined whether flies could discriminate between ACP and its deuterated counterpart d$_8$-ACP and whether the amount of concentration of the odorants can affect the behavioral preference action. Their result showed a preference of ACP against equal concentration (1:1) of d$_8$-ACP. But reducing the amount of concentration of d$_8$-ACP to $50\%$ gave no significant preferential outcome of the flies. The flies approximately equally filled the arms of the T-maze.

In a similar way, they tested whether the flies could discriminate between isotopomers 1-octanol (OCT, C$_8$H$_{18}$O) and its deuterated counterpart d$_{17}$-OCT (C$_8$D$_{17}$OH). The flies showed preference to the OCT with $>30\%$ excess flies in the arm. This discrimination against d$_{17}$-OCT was removed by reducing the concentration of d$_{17}$-OCT by $75\%$ (1:0.25). They observed that the arms were filled with approximate equal flies, signifying that the discrimination is based on odor perception.

To eliminate the doubt that flies might be using other features rather than olfactory sense to discriminate against a particular odorant, Franco and colleagues used anosmic mutants in their test. This should lead to an elimination of the differential response to the deuterated odorants. The anosmic \textit{Drosophila} Or83b$^1$ and Or83b$^2$ mutants showed no preferential discrimination against d$_8$-ACP and also against d$_{17}$-OCT. The mutants were distributed equally at the arms of the T-maze as would expect. This implies that the flies use olfaction alone to discriminate between a normal odorant and its deuterated counterpart \cite{franco2011reply}. Additionally, there should be a salient feature that results in the spontaneous discrimination against deuterated odorant from its normal counterpart.

Flies can be trained to recognize and avoid a particular odorant by the use of an electric foot-shock punishing stimulus \cite{tully1985classical}. In all the tests on 3 pairs of the odorants used by Franco and colleagues, flies continuously avoided shock-associated odor. This strengthens the idea that flies can distinguish between isotopomers. 

To examine the possibility that the spontaneous preferences seen on the flies are not as a result of impurities which the odorant might be contained even though the odorants are of high impurity-free, the flies were conditioned to generally avoid either deuterated or normal odorant from different unrelated odorants. The flies avoided all the unrelated deuterated odorants when trained to avoid a particular deuterated odorant. The same behavior is seen when the flies are conditioned to avoid particular natural hydrogenated odorants, they avoided all the unrelated normal odorants. This suggests that a salient feature such as the molecular vibrations of the odorant could be the feature the flies sense to distinguish isotopomers. If so, the flies use the modes most affected by deuteration (e.g., the C-H stretch) to generalize and distinguish deuterium from the normal odorants. The C-H stretch occurs approximately at 3,000 cm$^{-1}$ region but reduces to 2,200 cm$^{-1}$ region for deuterated odorants, C-D.

\begin{figure}
\includegraphics[scale=.65]{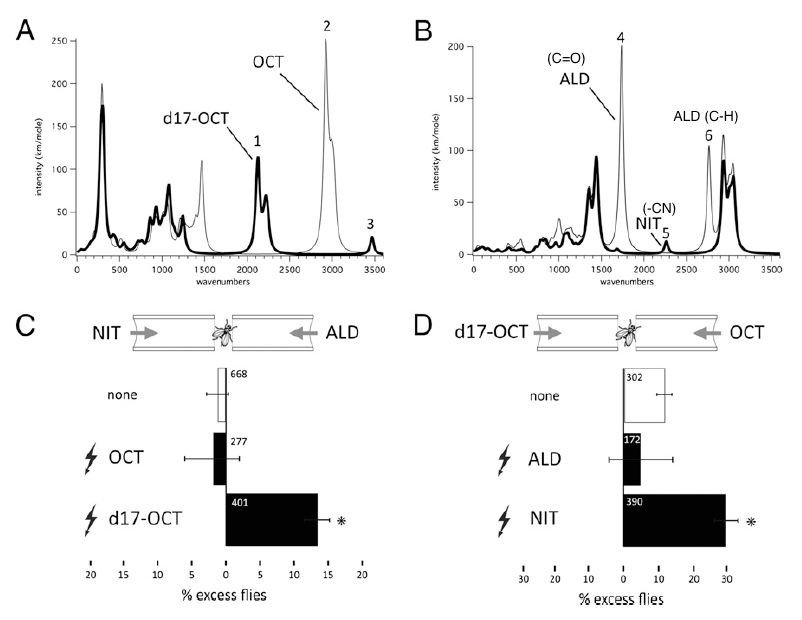}
\caption{A result of a T-maze experiment to show fruit flies use vibrational frequency in odor discrimination. FIG.(A) shows the computed vibrational spectra of OCT and d$_{17}$-OCT. As a result of deuteration, a reduction of the vibration stretch from 3,000cm$^{-1}$ (OCT) to 2,150cm$^{-1}$ (d$_{17}$-OCT) is obtained. FIG.(B) shows the computed vibration spectra for citronellal (ALD) and citronellyl nitrile (NIT). Both have almost the same vibrational frequency. ALD has two noticeable stretch frequencies; for C=O stretch at 1750cm$^{-1}$ and for C-H stretch at 2,765cm$^{-1}$, while NIT has a different stretch at 2,150cm$^{-1}$. FIGs. (C) and (B) show the odor discriminations by the flies when conditioned to avoid a particular odorant. The lightning symbol represents the electric foot shock used to condition the flies. Figure from \cite{franco2011reply}.}\label{fruitfly}
\end{figure}

Indeed, if the molecular vibrational frequency is the feature that determines the identification of odor, then flies should not be able to distinguish between odorants with the same known vibrational frequency and the same odor. Citronellyl nitrile (NIT, C$_{10}$H$_{17}$N) and citronellal (ALD, C$_{10}$H$_{18}$O) were used in the test since they possess similar odor characteristics to the human nose. Both have a lemongrass-like smell. Just like odor perception of human, flies showed no preference between the two odorants (FIG. \ref{fruitfly}C).
The IR spectra of both odorants in the fingerprint region are almost the same (FIG. \ref{fruitfly}B). They only differ in vibrations stretchs. Aldehyde has C=O stretch around 1,740cm$^{-1}$ and aldehydic C-H stretch around 2,765cm$^{-1}$. NIT and d$_{17}$-OCT share the same vibrational stretch at 2,150cm$^{-1}$. However, on conditioning of the flies to avoid d$_{17}$-OCT resulted in a discrimination against NIT. Likewise, conditioning the flies to avoid NIT (FIG. \ref{fruitfly}D) showed a discrimination against d$_{17}$-OCT. This suggests that flies detect the vibration of odorant's functional group and use it to discriminate among odorants. Some other researchers \cite[eg.,][]{bittner2012quantum,gronenberg2014honeybees} used similar method and similar odorant samples to show that just like mammals, insects can discriminate between isotopomers.


\subsection{Experiment refuting the vibrational theory of olfaction}
Inasmuch as so many authors accepted the vibrational theory as a plausible theory in olfaction, some still argue that it is not a viable theory for it fails to account for some differences in the smell of enantiomers. To prove the implausibility of vibration theory, Rajeev et al. \cite{muthyala2017testing} reported two different tests that confirmed the theory of vibration cannot be solely used to explain olfaction. They designed two different tests for groups of students for three years. These different groups had to test 1) the vibrational theory of olfaction using isotopomeric odorants and 2) the enantiomeric odorants. 

\begin{figure}

\includegraphics[scale=0.5]{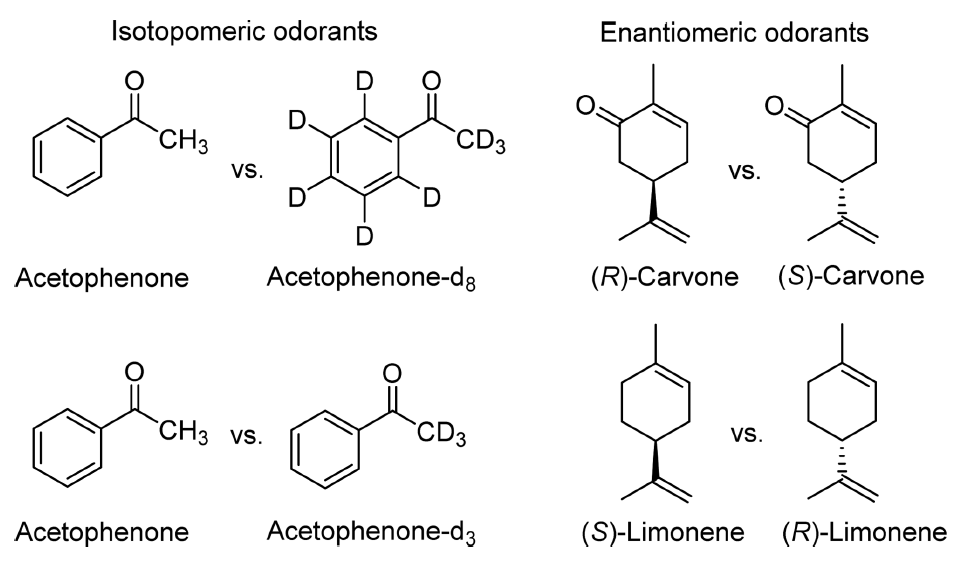}
\caption{The structures of the odorants used in the experiment. Figure from \cite{muthyala2017testing}.}
\label{odorant}
\end{figure}

A sample of acetophenone and aceptophenone-d$_3$ were used to test whether deuterated odorants and their nondeuterated counterparts have different odors. In the second test, a sample of (R)- and (S)-carvone, and (R)- and (S)-limonene (FIG. \ref{odorant}) were used to test whether enantiomeric odorants have the same odor since they ought to have the same bond vibrations. These tests can only validate the vibrational theory of olfaction:
\begin{itemize}
\item if the end of the experiment confirms that the deuterated compounds have a different odor compared to the nondeuterated compounds as predicted in literature.
\item if in the conclusion of the experiment, the enantiomers were confirmed to have the same odor since they have identical vibrational spectra.
\end{itemize}

\begin{figure*}
\includegraphics[scale=.5]{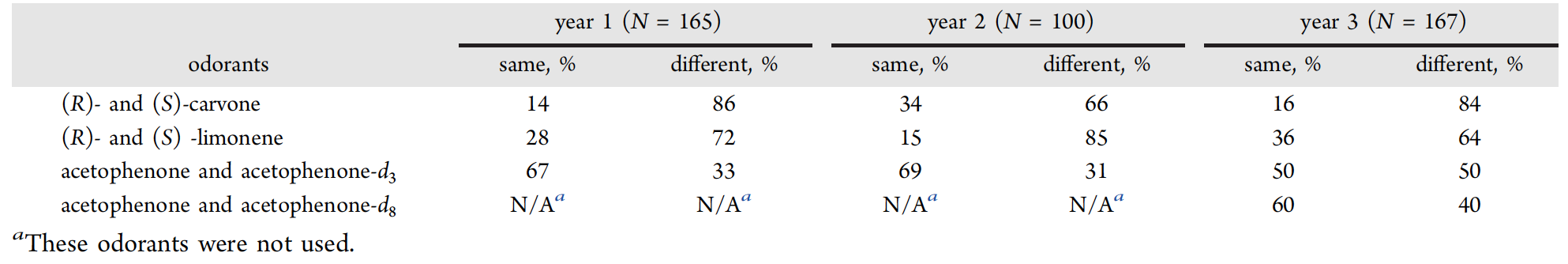}
\caption{A compilation of the result from the students on determining whether the odorants have the same smell or not \cite{muthyala2017testing}.}
\label{resultodorant}
\end{figure*}
Firstly, the students predicted and compared the stretching frequency of a C-D bond stretch and C-H bond by using the simple harmonic approximation to bond vibrations and applied Hooke's law. C-H bond was found to have a higher bond stretching frequency than the C-D bond confirming that deuterated compounds and its counterparts should have different vibrational frequencies. 

In the other hand, the measured IR spectra of the enantiomers were found to be identical while they have different odors. This does not necessarily generally conclude that all enantiomeric odorant have different odors even though the majority do \cite{bentley2006nose}. (R)-isomers of carvone are the most abundant compound in the essential oil from different species of mint thus, has a strong spearmint-leaves odor, while its mirror image (S)-carvone smells like caraway seeds \cite{murov1973odor,leitereg1971chemical}. The biosynthesis of carvone is by oxidation of limonene (which has a strong smell of oranges) \cite{karlberg1992air}. Limonene mainly occurs as the (R)-enantiomers with a strong smell of a fresh citrus orange, while (S)-limonene isomer has a hash, turpentine-like lemony smell \cite{friedman1971odor}.

It is noteworthy to point out that some other odorants can be used for this experiment. The choice for these odorants was influenced by their commercial availability, non-harmful nature and relatively affordable. Using a proper wafting technique, the students determined and recorded whether the odorants have the same smell or not. A compilation of the result is shown in table \ref{resultodorant}.

Comparing the percentage of the students reporting whether the enantiomers smell the same or different, it indicates that the acetophenone and acetophenon-d$_3$ have an identical odor, while the enantiomeric carvaones and limonenes have different odors for the first two years the tests were performed. In the third year, there was an equal percentage report of acetophenone and acetophenone-d$_3$ having a different smell and the same smell. But when acetophenone and acetophenone-d$_8$ were tested, the students reported they smell the same. The inconsistency of the result from first two years and the third year was attributed to be possibly due to the perireceptor events \cite{pelosi1996perireceptor} in the nasal mucus as reported by Brock et al. \cite{block2015implausibility}. The biochemical interactions between the odorants and the nasal mucus (in the vertebrates) or the sensillar lymph (in insects) can affect the final odor of the odorants. The presence of impurities in one of the commercial samples of acetophenone-d$_3$ was reported as possibly being the reason for different perceptions recorded by the students during the first two years and the third year. 

In conclusion, their perception result does not support the vibrational theory of olfaction as reported in literature because in the case of the isotopomers having the same odor, this is not in agreement with the vibrational theory stating that molecules having different vibration frequencies would smell differently. Also, if the shapes of the two enantiomers are not considered, then they should have the same vibration, thus the same smell, but that was not the case observed in the experiment. 

\section{Conclusion and summary}
In this paper, we have studied the theories of olfaction. The studied theories include the `lock and key' model, the odotope model, vibrational theory by Dyson, Swipe-card model by Stoneham, Turin theory of inelastic electron tunneling spectroscopy (IETS) and chiral recognition models. Among all these, Turin's model has gained more popularity. There are some experimental evidence that support the idea of detection of smell to be a function of the molecular frequency, while some experiments have showcased that vibrational theory of olfaction is implausible and as such, should not be regarded as an explanation of odor detection and disquisition. 

It is still an unsolved question whether recognition of smell is either solely by shape theory or the molecular vibrations of the odorant or a combination of the two \cite{gane2013molecular}. However, since olfactory receptors belong to the family of class A G protein-coupled receptors, and since proteins are chiral, then shape theory provides an insufficient explanation to why most pair of enantiomers has the same odor.

The vibrational theory has faced a lot of antagonism over its inability to reconcile the fact that some enantiomers have different smell even when they have the same vibrational spectrum.

An experiment by Franco and colleagues has suggested that fruit flies use the molecular vibration of an odorant for its preferential selection and discrimination. In contrast, the result of the experiment by Rajeev and colleagues refutes the vibrational theory of olfaction since they obtained the same smell for isotopomers used in their test, while the vibrational theory states that molecules with different vibrational frequencies should have different smells. Also, enantiomers with the same vibration frequency should smell the same because if shape is neglected, bond is the only thing that affects vibration. But in their own case, they observed different smells for their enantiomers.

Although the theories of olfaction are faced with some challenges and some unanswered questions (like how the shape of the olfactory receptor looks), I believe olfaction has a connection with a quantum mechanical principle, thus a focus on a biological mechanism obtained from a principle in quantum mechanics  should result to a permanent solution to the unanswered questions in the field of quantum biology.

\section{\label{sec:citeref}References}





\nocite{*}

\bibliography{Emeka_phy771}

\end{document}